\documentclass[12pt]{iopart}

\newcommand{\td} {\mathrm{d}}
\usepackage{color}
\newcommand{\bm}[1]{\mbox{\boldmath{$#1$}}}

\begin{document}

\title{Quasi-local energy and the choice of reference}

\author{Jian-Liang Liu$^{1}$,  Chiang-Mei Chen$^{1}$, and James M Nester$^{1,2}$}

%%%jmn I put JL Liu as the first author.

\address{$^{1}$Department of Physics \& Center for Mathematics and Theoretical Physics, National Central University, Chungli 320, Taiwan}
\address{$^{2}$Graduate Institute of Astronomy, National Central University, Chungli 320, Taiwan}

%\ead{chiangmeichen@gmail.com, tendauliang@gmail.com, and jmnester@gmail.com}
%%%jmn It is better to use cmchen@phy.ncu.edu.tw and nester@phy.ncu.edu.tw
\ead{liujl@phy.ncu.edu.tw, cmchen@phy.ncu.edu.tw, and nester@phy.ncu.edu.tw}
%%%Liu_110502%%% change to the NCU email %

\begin{abstract}
A quasi-local energy for Einstein's general relativity is defined
by the value of the preferred boundary term in the covariant Hamiltonian formalism.
The boundary term depends upon a choice of reference and a time-like displacement
vector field (which can be associated with an observer) on the boundary of the region.
Here we analyze the spherical symmetric cases.
For the obvious analytic choice of reference based on the metric
components, we find that this technique gives the same quasi-local energy values using several standard coordinate systems
and yet can give different values in some other coordinate systems.
For the homogeneous-isotropic cosmologies, the energy can be non-positive, and one case which is actually flat space has a negative energy.
As an alternative, we introduce a way to determine the choice of both the reference and displacement by extremizing the energy.  This procedure gives the same value for the energy in different coordinate systems for the Schwarzschild space, and a non-negative value for the cosmological models, with zero energy for the dynamic cosmology which is actually Minkowski space.  The timelike displacement vector comes out to be the dual mean curvature vector of the two-boundary.
\end{abstract}

\section{Introduction}

The identification of a good expression which well describes the energy for gravitating systems (and really this means
 for all physical systems) still remains an outstanding issue.  One consequence of the equivalence principle is that
  there is no well defined (i.e., covariant) {\em local\/} description of the energy-momentum for gravitating systems
   (see, e.g., the discussion in \S 20.4 of~\cite{Misner:1974qy}).  The modern idea is that properly energy is
    {\em quasi-local\/} (i.e., is associated with a closed 2-surface, for a comprehensive review
    see~\cite{Szabados:2004vb}).  One approach is to regard energy as the value of the generator of dynamical
    changes with time, the Hamiltonian.  Here we consider in particular the covariant Hamiltonian formalism~
    \cite{Chen:1994qg,Chen:1998aw,Chen:2000xw,Chen:2005hwa}.  Within that approach a certain preferred Hamiltonian
    boundary term was identified~\cite{Chen:2005hwa}. The quasi-local energy is given by the value of this boundary
    term with a suitable choice of time evolution vector field on the closed 2-surface.
    In addition to the spacetime displacement vector field and the dynamical variables,
    this boundary term also depends in general on a choice of certain reference values for
    the dynamical variables (which
    represent %%%Liu_110830%%%
    the ground state with vanishing quasi-local quantities). Unlike the case for other fields, the reference values for gravity theories based on dynamic geometry (the metric and connection) cannot be taken to have trivial values.  Essentially this is because the natural ground
     state for dynamic geometry, Minkowski space, has a {\em non-vanishing} metric.
  Hence the choice of reference for such theories %%%Liu_110704%%%
  in general, and thus %%%Liu_110720
  in particular %%%Liu_110720
  for general relativity, %%%Liu_110704%%%
 necessarily requires some suitable way to select an appropriate Minkowski geometry at the points of the closed 2-surface.
 Presently this is a quite active research topic~\cite{Wang:2008jy,WaYa2008,Liu:2003bx}. Here, following~\cite{Liuthes},
  we consider this problem for the most important special case: spherically symmetric solutions (more specifically,
  Schwarzschild and the homogeneous isotropic cosmologies) to Einstein's gravity theory, General Relativity (GR).
  We consider two approaches.  The first %%%Liu_110704%%%
 we name %%%Liu_110704%%%
 {\em analytic}---essentially one takes the limit of the physical metric components in some coordinate system when the physical parameters take on
  trivial values.  This type of approach goes back to~\cite{Chen:1998aw,CMthesis},
  %%%Liu_110719
    and shows that the quasi-local energy is reference and displacement vector dependent.%%%jmn110722
     That naturally %%%Liu_110720
 raises %%%jmn110722  leads to %%%Liu_110720
    a question: is there a minimum (maximum) value among all these available choices? Which leads to the %%%Liu_110720
    second approach: find %%%Liu_110720
    the optimal reference via a variational principle extremizing the energy.
    Here we show that the analytic approach can give %%%jmn110722 lead to
    the same standard quasi-local energy value for several choices of the spatial coordinate
    system and yet will lead to different energy values for different time coordinates.  The value of the obtained quasi-local energy is not
    necessarily
    non-negative.  %%%Liu_110720
    On the other hand the energy extremization always gives %%%jmn110722 leads to
    a coordinate frame independent quasi-local energy value which is, moreover, non-negative and vanishing only for Minkowski space.
 %%%Liu_110719

The outline of this work is as follows.  In Section 2, we briefly introduce the covariant Hamiltonian approach, which leads to the Hamiltonian boundary
 term that gives the quasi-local energy.  Section 3 concerns the analytic approach to choosing the reference: $3.1$ includes the analysis of the
  Schwarzschild geometry in three  different spatial coordinate systems and the Eddington-Finkelstein and Painlev\'e-Gullstrand time slicings;
  the homogeneous-isotropic cosmology metrics are considered in Section 3.2. In section $4$, we use the method %%%Liu_110718%%%
 of extremizing the quasi-local energy to determine the choice of reference and also the displacement vector. Section 5 is the conclusion.

\section{The covariant Hamiltonian approach}

 Our approach to quasi-local energy is via the covariant Hamiltonian formalism, which has been described in detail in a series of
 works~\cite{Chen:1994qg,Chen:1998aw,Chen:2000xw,Chen:2005hwa}. The construction of the energy expression is briefly outlined here.
 It begins from a {\em first order Lagrangian} 4-form for some $k$-form field.
%%%Liu_110718 (more generally one may have a collection of several form fields which can carry also some suppressed tensor indicies):
 \begin{equation}
 {\cal L}=\td\varphi\wedge p-\Lambda(\varphi,p).\label{L1}
 \end{equation}
 The variation has the form
 \begin{equation}
 \delta{\cal L}=\td (\delta\varphi\wedge p)+\delta\varphi\wedge \frac{\delta
\mathcal{L}}{\delta\varphi}+\frac{\delta
\mathcal{L}}{\delta p}\wedge\delta p.\label{deltaL1}
 \end{equation}
 Hamilton's principle applied to the action obtained by integrating the first order Lagrangian over a region leads to the pair of {\em first-order} field equations:
 \begin{equation}
 0=\frac{\delta
\mathcal{L}}{\delta\varphi}:=-(-1)^k\td p-\partial_\varphi\Lambda,\quad 0=\frac{\delta
\mathcal{L}}{\delta p}:=\td\varphi-\partial_p\Lambda.\label{1stordFE}%%%Liu_110429%%% a sign error %
 \end{equation}
 From infinitesimal diffeomorphism invariance (generated by a vector field $\textbf{N}$) for ${\cal L}$ it follows that (\ref{deltaL1}) must become an identity under the replacement $\delta\to \pounds_{\textbf{\tiny{N}}}$ (where $\pounds_{\textbf{\tiny{N}}}=\td\iota_{\textbf{\tiny{N}}}+\iota_{\textbf{\tiny{N}}}\td$ on the components of form fields):
 \begin{equation}
 \td\iota_{\textbf{\tiny{N}}}{\cal L}\equiv \pounds_{\textbf{\tiny{N}}}{\cal L}\equiv\td (\pounds_{\textbf{\tiny{N}}}\varphi\wedge p)+\pounds_{\textbf{\tiny{N}}}\varphi\wedge \frac{\delta
\mathcal{L}}{\delta\varphi}+\frac{\delta
\mathcal{L}}{\delta p}\wedge\pounds_{\textbf{\tiny{N}}} p.
 \end{equation}
 From this identity it follows that the 3-form
 \begin{equation}
 {\cal H}(\textbf{N}):=\pounds_{\textbf{\tiny{N}}}\varphi\wedge p-\iota_{\textbf{\tiny{N}}}{\cal L}\label{calH}
 \end{equation}
 satisfies the (a particular case of Noether's first theorem) identity
 \begin{equation}-\td{\cal H}(\textbf{N})\equiv\pounds_{\textbf{\tiny{N}}}\varphi\wedge \frac{\delta
\mathcal{L}}{\delta\varphi}+\frac{\delta
\mathcal{L}}{\delta\varphi}\wedge\pounds_{\textbf{\tiny{N}}} p.\label{dH}\end{equation}%%%Liu_110429%%% a sign error %
 Hence (when the field equations are satisfied) the 3-form ${\cal H}(\textbf{N})$ is a conserved current.
 From an expansion of its definition (\ref{calH}) it can be seen to have the form
 ${\cal H}(\textbf{N})=N^\mu{\cal H_\mu}+\td{\cal B}(\textbf{N})$.  Inserting this expansion into the identity
 (\ref{dH}) gives %%%jmn110722 leads to
 an identity with terms proportional to  $\td N^\mu$ and $N^\mu$, which should hold at any point for arbitrary values of these variables.  From the terms proportional to $\td N^\mu$ one learns that for any diffeomorphically invariant Lagrangian, $\mathcal{H}_{\mu}=-\iota_{\mu}\varphi\wedge\frac{\delta
\mathcal{L}}{\delta\varphi}+(-1)^{k}\frac{\delta \mathcal{L}}{\delta p}\wedge\iota_{\mu} p$ (this is a special application of Noether's second theorem), and thus ${\cal H}_\mu$ {\em vanishes} when the field equations are satisfied.  Consequently, the value of the conserved current associated with a spatial region $\Sigma$ and vector field $\textbf{N}$,
\begin{equation}
E(\textbf{N},\Sigma):=\int_{\Sigma}\mathcal{H}(\textbf{N})=\int_{\Sigma}N^{\mu}\mathcal{H}_{\mu}+\oint_{\partial\Sigma}\mathcal{B}(\textbf{N})
=\oint_{\partial\Sigma}\mathcal{B}(\textbf{N})\label{energy}
\end{equation}
really depends only on the boundary term, and hence is actually {\em quasi-local}.
%%%Liu_110718\begin{equation}
%E(\textbf{N},\partial\Sigma)=\oint_{\partial\Sigma}\mathcal{B}(\textbf{N}).\label{energy}
%%%Liu_110718\end{equation}
The physical interpretation of this value is that for suitable displacements on the boundary it represents the components of the {\em quasi-local energy-momentum}.
However it must be noted that (just like other Noether conserved currents) one can add any total differential to (\ref{calH}).
The resulting 3-form would still be a conserved current, but it would define a different conserved value.  By this process one can adjust
the boundary expression ${\cal B}(\textbf{N})$ and thereby the associated conserved energy-momentum to have almost any value.  Fortunately
the 3-form (\ref{calH}) has another important role which brings the freedom  in the choice of the boundary term ${\cal B}$, and hence the value
$E(\textbf{N},\partial \Sigma)$, under physical control.  The 3-form ${\cal H}(\textbf{N})$ is not only the conserved current associated with an
infinitesimal diffeomorphism it is also actually the {\em Hamiltonian density}; i.e.,  the Hamiltonian $H(\textbf{N},\Sigma)$ %%%Liu_110720
that generates the dynamical evolution of the variables along a timelike vector field $\textbf{N}$ is just (\ref{energy}).
%%%Liu_110718\begin{equation}
%\textbf{H}(\textbf{N},\Sigma)=\int_{\Sigma}\mathcal{H}(\textbf{N})=\int_{\Sigma}N^{\mu}\mathcal{H}_{\mu}+\oint_{\partial\Sigma}\mathcal{B}(\textbf{N}).
%\label{Hamiltonian-3}
%%%Liu_110718\end{equation}
The variation of this Hamiltonian has the form
\begin{eqnarray}
\delta H(\textbf{N},\Sigma)&=&\int_{\Sigma}(\delta\varphi\wedge\pounds_{{\textbf{\tiny N}}}
p+\pounds_{{\textbf{\tiny N}}}\varphi\wedge \delta
p)+\oint_{\partial\Sigma}\mathcal{C}(\textbf{N}),\label{vari-quasilocal-onshell}
\end{eqnarray}
yielding %%%jmn110722 leading to
the Hamilton equations
$\pounds_{{\textbf{\tiny N}}}\varphi={\delta\mathcal{H}( \textbf{N})}/{\delta
p}$, $\pounds_{{\textbf{\tiny N}}}p=-{\delta\mathcal{H}(\textbf{N})}/{\delta\varphi}$. %%%jmn110722
% $\pounds_{{\textbf{\tiny N}}}\varphi=\frac{\delta\mathcal{H}( \textbf{N})}{\delta
% p}$, $\pounds_{{\textbf{\tiny N}}}p=-\frac{\delta\mathcal{H}(\textbf{ N})}{\delta\varphi}$.%%%jmn110722
%%%Liu_110718\begin{equation}
%\pounds_{{\textbf{\tiny N}}}\varphi=\frac{\delta\mathcal{H}(\textbf{N})}{\delta
%p},\quad\pounds_{{\textbf{\tiny N}}}p=-\frac{\delta\mathcal{H}(\textbf{N})}{\delta\varphi}.
%\end{equation}
%%%Liu_110718
 The key point is that requiring \emph{functional differentiability} of the Hamiltonian (i.e., the vanishing of the boundary term in the variation of the Hamiltonian), determines the \emph{boundary conditions} built into the Hamiltonian.
Hence one should thus choose the particular form of the Hamiltonian boundary term ${\cal B}$ that
gives %%%jmn110722 leads to
the desired type of boundary condition for the dynamical variables (e.g., Dirichlet or Neumann) which is suitable for the physical problem.

%%%Liu_110718
The boundary term $\mathcal{C}(\textbf{N})$ in the variation of the Hamiltonian (\ref{vari-quasilocal-onshell}), will not have vanishing limiting
value at infinity with the usual field fall offs~\cite{Regge:1974zd,BeOm1987}---unless one adjusts %%%Liu_110429%%% cancle one more adjusts %
{\em by hand} the total differential (i.e., the boundary term) in the Hamiltonian.

Investigations led to several explicit alternative boundary term expressions.
In general these expressions require, along with the dynamical variables on the boundary certain {\em non-dynamical} reference values, that represent the ground state of the physical system.  Whereas it is generally possible, and indeed appropriate, to choose trivial (i.e., vanishing) magnitudes for these reference values for all the other physical fields, that cannot be done for dynamic geometry gravity theories, simply because the ground state of the metric is not a vanishing value but rather the non-vanishing Minkowski metric.  Moreover, in a general coordinate system, the Minkowski connection also has non-vanishing components.

 Among the possible boundary terms corresponding to various boundary conditions,
  a certain preferred boundary term for the covariant Hamiltonian for Einstein's GR was identified which should be suitable for most applications. It corresponds to holding the metric fixed on the boundary. This choice has the virtue of directly giving not only the ADM quantities at spatial infinity but also the Bondi energy and flux at null infinity, and moreover under certain conditions it will give a {\em positive} quasi-local energy.  Our preferred boundary term for GR is given by
\begin{equation}
\mathcal{B}(\textbf{N})=\frac{1}{2\kappa}(\Delta{\Gamma^{\mu}}_{\nu}
\wedge\iota_{{\textbf{\tiny N}}}{\eta_{\mu}}^{\nu}
+\bar{D}^{[\nu}\bar{N}^{\mu]}\Delta\eta_{\mu\nu}),\label{cov-bound-GR01}
\end{equation}
where %%%Liu_110704%%%
$\bar{D}^{\nu}\bar{N}^{\mu}=\bar{g}^{\nu\lambda}\bar{D}_{\lambda}\bar{N}^{\mu}
=\bar{g}^{\nu\lambda}(\partial_{\lambda}\bar{N}^{\mu}+\overline{\Gamma}^{\mu}{}_{\lambda\gamma}\bar{N}^{\gamma})$,
%%%Liu_110704%%%
 $\Delta{\Gamma^{\mu}}_{\nu}={\Gamma^{\mu}}_{\nu}-{\overline{\Gamma}^{\mu}}_{\nu}$ is the difference of the connection one-form between the dynamic space and the reference space,
with a bar referring to the reference objects, %%%jmn110501%%%Liu_110720
 $\eta_{\mu\nu}=\frac{1}{2}\epsilon_{\mu\nu\alpha\beta}\vartheta^{\alpha}\wedge\vartheta^{\beta}$, %%%Liu_110720
 and $\vartheta^{\mu}$ is the orthonormal coframe.

For a given dynamical region and given dynamical fields, the value of this boundary expression with
suitably selected vector fields can be used to determine values for the quasi-local energy-momentum.
However, to find the specific values for these quasi-local quantities, one still needs to explicitly
select the reference configuration and the appropriate displacement vector field.
%A natural choice for the reference is Minkowski space.  But one must determine exactly which
%Minkowski space should be used.  In the following sections we explore two approaches to achieving this.
%An intuitive choice for the reference is induced by the dynamic space, which is the approach in the next
%section. Later we make another approach by keeping the Minkowski metric with a \emph{specific} connection as the reference space.
%%%Liu_110429%%% change the argument a little bit %
A natural choice for the reference is one with a Minkowski metric.  But one must determine exactly which %%%Liu_110718
Minkowski space should be used.  In the following sections we explore two approaches to achieving this.
%%%jmn110501

\section{Choice of reference: the analytic approach}
The boundary expression (\ref{cov-bound-GR01}) includes the reference values for the dynamical fields,
but gives no restriction as to what the reference should be. In general, the reference space could be
any four dimensional manifold endowed with a Lorentz signature metric tensor and a connection.
%Here we will, as usual, take the reference space to be a Minkowski space.%
%%%Liu_110429%%%
%Here we take the reference space to be the Minkowski metric and a connection induced by the dynamic one. %%%Liu_110429%%%
Here we take the reference space to have a Minkowski metric.  %%%jmn110501
 The quasi-local energy (\ref{energy}) is the value determined by the \emph{boundary integral} with a certain time displacement. The boundary term (\ref{cov-bound-GR01}) includes the reference variables for (certain projected components of) the four-dimensional metric and connection on the two-boundary. A reasonable requirement for choosing the reference is isometric embedding of the two-boundary into the chosen reference space.
 %%%Liu_110718
 %Since the Hamiltonian value is defined \emph{quasi-locally}, it is appropriate for the two-surfaces to be taken to be the same (in the sense of
 %\emph{isometric}) in both the dynamic spacetime and the reference space.%%%Liu_110718
 Without additional conditions, however, the embedding is not unique.

Here we consider just spherical symmetric spacetimes. This is an important yet relatively simple special case for finding %%%Liu_110720
an %%%Liu_110720
isometric embedding easily, and also it simplifies the Hamiltonian boundary expression (\ref{cov-bound-GR01}), so that only the first term $\Delta{\Gamma^{\mu}}_{\nu}\wedge\iota_{{\textbf{\tiny N}}}{\eta_{\mu}}^{\nu}$ contributes.

\subsection{Schwarzschild geometry}

For the static, spherically symmetric Schwarzschild metric, we will consider five different representations related to different time and spatial coordinates: the standard $\{t,r\}$ where $r$ is the {\em area coordinate},  the {\em isotropic spherical} $\{t,R\}$ and {\em isotropic Cartesian} $\{t,x,y,z\}$ coordinates, and the Eddington-Finkelstein $\{\tilde{t},r\}$ and Painlev\'e-Gullstrand $\{\tau,r\}$ coordinates. In each case the choice of reference is obtained {\em analytically} by taking $m=0$ in the metric and connection coefficients of the dynamic spacetime in the particular coordinates chosen.  In general it can be expected that the resulting quasi-local energy value will depend on the choice of coordinates.

Let us first illustrate the procedure in detail using the standard area coordinate.
\subsubsection{Standard Schwarzschild.}
The standard form of the Schwarzschild metric is
\begin{equation}
\td s^{2}=-\left(1-\frac{2m}{r}\right)\td t^{2}+\left(1-\frac{2m}{r}\right)^{-1}\td r^{2}+r^{2}\td\theta^{2}+r^{2}\sin^{2}\theta\,\td\varphi^{2}.\label{sch-met}
\end{equation}
The radial coordinate is determined geometrically in terms of the area of a 2-sphere: $A=4\pi r^2$.
Take the orthonormal coframe to be
\begin{equation}
\vartheta^{0}=\sqrt{1-\frac{2m}{r}}\td t,\quad\vartheta^{1}=\frac{1}{\sqrt{1-\frac{2m}{r}}}\td r,\quad\vartheta^{2}=r\td \theta,\quad\vartheta^{3}=r\sin\theta\,\td \varphi.\label{Standard Sch coframe}
\end{equation}
The Levi-Civita connection one-form coefficients are obtained using the torsion free condition, $\td\vartheta^{\mu}+{\Gamma^{\mu}}_{\nu}\wedge\vartheta^{\nu}=0$.  Due to the metric compatibility condition, the orthonormal frame connection coefficients are {\em anti-symmetric}.  The independent connection coefficients are
\begin{eqnarray}
&&{\Gamma^{1}}_{2}=-\sqrt{1-\frac{2m}{r}}\td\theta,\quad
{\Gamma^{1}}_{3}=-\sqrt{1-\frac{2m}{r}}\sin\theta\,
\td\varphi, %\nonumber\\
\quad {\Gamma^{2}}_{3}=-\cos\theta \td\varphi\nonumber\\
&&{\Gamma^{0}}_{1}=\frac{m}{r^{2}}\td t,\quad
{\Gamma^{0}}_{2}=0,\quad{\Gamma^{0}}_{3}=0.\label{connection-Sch}
\end{eqnarray}
Take $m=0$ in (\ref{Standard Sch coframe}) and (\ref{connection-Sch}) to obtain the reference geometry components. Then the non-vanishing differences of the connection components in (\ref{cov-bound-GR01}) become
\begin{eqnarray}
&&\Delta{\Gamma^{0}}_{1}=\frac{m}{r^{2}}\textrm{d}t,\nonumber\\
&&\Delta{\Gamma^{1}}_{2}=\left(1-\sqrt{1-\frac{2m}{r}}\right)\textrm{d}\theta,\quad
\Delta{\Gamma^{1}}_{3}=\left(1-\sqrt{1-\frac{2m}{r}}\right)\sin\theta\,
\textrm{d}\varphi. \label{connection-Sch-r-difference}
\end{eqnarray}
Note that the term $\Delta\eta_{\mu\nu}$ vanishes in the boundary integral for this reference choice, so the boundary expression (\ref{cov-bound-GR01}) reduces to just the first term.

Another important role in the boundary expression is played by the displacement vector. We assume it is a \emph{future time-like} vector field. Suppose \textbf{N} is normal to the two-surface (which we choose here to be the constant $t$, $r$ sphere,  with $\textbf{e}_{2}$, $\textbf{e}_{3}$ being the two-surface tangent vectors), then the factor $\iota_{\textbf{{\tiny N}}}{\eta_{\mu}}^{\nu}$ is obtained using $\textbf{N}=N^{0}\textbf{e}_{0}+N^{1}\textbf{e}_{1}$:
\begin{eqnarray*}
&&\iota_{\textbf{{\tiny N}}}{\eta_{0}}^{1}=0,\quad\iota_{\textbf{{\tiny N}}}{\eta_{0}}^{2}=-r\sin\theta N^{1}\td\varphi,\quad\iota_{\textbf{{\tiny N}}}{\eta_{0}}^{3}=rN^{1}\td\theta,\\
&&\iota_{\textbf{{\tiny N}}}{\eta_{1}}^{2}=r\sin\theta N^{0}\td\varphi,\quad\iota_{\textbf{{\tiny N}}}{\eta_{1}}^{3}=-rN^{0}\td\theta,\\
&&\iota_{\textbf{{\tiny N}}}{\eta_{2}}^{3}=\frac{N^{0}}{\sqrt{1-2m/r}}\td r-\sqrt{1-2m/r}N^{1}\td t.
\end{eqnarray*}
Only the purely angular components of the quasi-local boundary term will contribute to the integral over the 2-sphere $S$:
\begin{eqnarray}
2\kappa\mathcal{B}(\textbf{N})&=&\Delta{\Gamma^{\mu}}_{\nu}
\wedge\iota_{{\textbf{\tiny N}}}{\eta_{\mu}}^{\nu} \nonumber\\
&=&
2[\Delta \Gamma^1{}_2\wedge \iota_N\eta_1{}^2+\Delta \Gamma^1{}_3\wedge \iota_N\eta_1{}^3]\nonumber\\
&=&4(1-\sqrt{1-2m/r})N^0\, r\sin\theta\, \td\theta\wedge\td\varphi.%%%Liu_110429%%%change \phi to \varphi%%%Liu_110429%%%
\end{eqnarray}
The quasi-local energy obtained from the integral over the 2-sphere boundary at constant $t,r$ then comes out to be
\begin{equation}
E_{\rm S}(\textbf{N})=r(1-\sqrt{1-2m/r})N^{0}.
\label{quasi-energy-Sch}
\end{equation}
One possible choice of the displacement is the time-like Killing vector of the reference $\textbf{N}=\partial_{t}=\sqrt{1-2m/r}\,\textbf{e}_{0}$, yielding
\begin{equation}
E_{\rm S}(\partial_{t})=r\sqrt{1-2m/r}(1-\sqrt{1-2m/r}),\label{energy-Sch-ddt01}
\end{equation}
with the horizon and asymptotic limits
\begin{eqnarray}
&&E_{\rm S}(\partial_t)_{r\rightarrow\infty}\approx m\left(1-\frac{m}{2r}\right)\to m,\quad
%%%Liu_110718\label{energy-Sch-ddt-infinity}\\
E_{\rm S}(\partial_t)_{r=2m}=0.\label{energy-Sch-ddt-2M}
\end{eqnarray}
On the other hand, for the choice of $\textbf{N}=\textbf{e}_{0}$ (this is the \emph{unit dual mean curvature vector} of the two-surface), the quasi-local energy value comes out to be
\begin{equation}
E_{\rm S}(\textbf{e}_0)=r(1-\sqrt{1-2m/r}),\label{energy-Sch-e0}
\end{equation}
with the horizon and asymptotic limits
\begin{eqnarray}
&&E_{\rm S}(\textbf{e}_0)_{r\rightarrow\infty}\approx m\left(1+\frac{m}{2r}\right)\to m,\quad
%%%Liu_110718\label{energy-Sch-e0-infinity}\\
E_{\rm S}(\textbf{e}_0)_{r=2m}=2m.\label{energy-Sch-e0-2M}
\end{eqnarray}
The value (\ref{energy-Sch-e0}) is the famous result first found by Brown and York~\cite{Brown:1992br}.
It turns out that several approaches yield this same value, %%%Liu_110720
so we will refer to it as the {\em standard} value.
From an examination of these two results we have noticed a curious fact:
\begin{equation}
E_{\rm S}\left(\frac12(\partial_t+\textbf{e}_0)\right)=m.
\end{equation}
We do not know whether this has any significance.

The calculations for several other metric expressions of interest to us here follow a similar common procedure, to avoid unnecessary repetition we have done the general calculation in the {\bf Appendix};  we briefly report in the following the specific results.

%------------------------------------------------------------------------------------------------------------------------------------------------------
%additional part for Isotropic Schwarzschild metric
%------------------------------------------------------------------------------------------------------------------------------------------------------

\subsubsection{Isotropic Schwarzschild} The spherical isotropic Schwarzschild metric can be obtained from (\ref{sch-met}) using the coordinate transformation $r=R(1+\frac{m}{2R})^{2}$:
\begin{equation}
\td s^{2}=-\frac{(1-\frac{m}{2R})^{2}}{(1+\frac{m}{2R})^{2}}\td t^{2}+\left(1+\frac{m}{2R}\right)^{4}(\td R^{2}+R^{2}\td\theta^{2}+R^{2}\sin^{2}\theta\,\td\varphi^{2}).
\end{equation}
We choose the coframe
\begin{eqnarray}
&&\vartheta^{0}=\frac{1-m/2R}{1+m/2R}\td t,\quad\vartheta^{1}=(1+m/2R)^{2}\td R,\nonumber\\
&&\vartheta^{2}=(1+m/2R)^{2}R\td\theta,\quad\vartheta^{3}=(1+m/2R)^{2}R\sin\theta\,\td\varphi.\label{Schw-sph-iso-coframe}
\end{eqnarray}
 Then we worked out the corresponding Levi-Civita connection ${\Gamma^{\mu}}_{\nu}$ and took $m=0$ to get the reference values (${\overline{\Gamma}^{\mu}}_{\nu}$ and $\bar{\vartheta}^{\mu}$). The displacement vector in the normal space of the two-surface has the general form $\textbf{N}=N^{0}\textbf{e}_{0}+N^{1}\textbf{e}_{1}$. Only the angular components of the quasi-local boundary term contribute.

This procedure leads to the quasi-local energy for the spherical isotropic metric.
The value can be calculated from expression (\ref{A:Eschwi}) in the appendix, which gives
\begin{equation}
E_{\rm SI}({\bf N})=N^{0}E_{\rm SI}({\bf e}_0)=m(1+m/2R)N^{0}.\label{quasi-energy-isoSch}
\end{equation}
%%%jmn110501
%
%\begin{equation}
%E_{\rm SI}({\bf N})=m(1+m/2R)N^{0}.\label{quasi-energy-isoSch}
%\end{equation}
%%%Liu_110429%%%
%Note that $\Gamma^{\tau}{}_{\theta}$ and $\Gamma^{\tau}{}_{\varphi}$ vanish such that
%$E_{\rm SI}({\bf N})=N^{0}E_{\rm SI}({\bf e}_{\perp})$, where $\textbf{e}_{\perp}=\textbf{e}_{0}$.
%%%Liu_110429%%%
If we choose $\textbf{N}=\partial_{t}$ (the timelike Killing field of the reference), which means $N^{0}=\frac{1-m/2R}{1+m/2R}$, then
\begin{eqnarray}
E_{\rm SI}(\partial_t)=m(1-m/2R),\quad%%%Liu_110718
E_{\rm SI}(\partial_t)_{R\rightarrow m/2}=0,\quad%%%Liu_110718
E_{\rm SI}(\partial_t)_{R\rightarrow\infty}=m.
\end{eqnarray}
%We can see that the energy is \emph{negative} when $R<m/2$, but the isotropic coordinate does not cover the inner horizon region for the standard Schwarzschild coordinate $R>0$. So that here we should consider the region only for $\rho>m/2$.
Another choice for the evolution vector is $\textbf{N}=\textbf{e}_{0}$, which gives
\begin{equation}
E_{\rm SI}(\textbf{e}_0)=m(1+m/2R),\label{quasi-energy-isoSch1}
\end{equation}
with the limits
\begin{eqnarray}
E_{\rm SI}(\textbf{e}_{0})_{R=m/2}=2m,\quad%%%Liu_110718
E_{\rm SI}(\textbf{e}_{0})_{R\rightarrow\infty}=m.%%%Liu_110718
\end{eqnarray}
The value (\ref{quasi-energy-isoSch1}) is %%%Liu_110704%%%actually %%%Liu_110704%%%
just %%%Liu_110720
the standard result (\ref{energy-Sch-e0}), after taking into account the transformation $r=R(1+m/2R)^{2}$.

  Furthermore, we considered another metric form which is also isotropic, but in the Cartesian coordinate system $x^{\mu}=\{t,x,y,z\}$. This is an important check on our techniques, since for this representation the reference connection vanishes when we take $m=0$ in the dynamic connection.  With $x=R\sin\theta\cos\varphi$, $y=R\sin\theta\sin\varphi$, $z=R\cos\theta$, then $R^{2}=x^{2}+y^{2}+z^{2}$. The metric then has the form
\begin{equation}
\td s^{2}=-N^{2}\td t^{2}+\Phi^{2}(\td x^{2}+\td y^{2}+\td z^{2}),\label{Iso_Sch_Cart_metric}
\end{equation}
where $N=\frac{1-m/2R}{1+m/2R}$, $\Phi=(1+m/2R)^{2}$. We choose the obvious coframe:
\begin{equation}
\vartheta^{0}=N\td t,\quad\vartheta^{i}=\Phi\td x^{i},\quad x^{i}=\{x,y,z\}.\label{Schw-cart-iso}
\end{equation}
Suppose that $N^{0}$ depends on $(t,R)$ only. Then, from the calculation given in more detail in the appendix, we find for the Cartesian coordinate isotropic Schwarzschild metric {\em exactly the same quasi-local energy result},
(\ref{quasi-energy-isoSch1}), as was found using  spherical coordinates
for the isotropic Schwarzschild metric.
%\begin{eqnarray}
%E=\frac{m(1+m/2\rho)}{2\kappa\rho^{3}}N^{0}\oint\epsilon_{ijk}x^{i}\td x^{j}\wedge\td x^{k}
%=m(1+2m/\rho)N^{0}.
%\end{eqnarray}
%Compare with (\ref{quasi-energy-isoSch}), the quasi-local energies of the isotropic Schwarzschild in both the spherical and Cartesian coordinate systems are identical.

Here we have calculated the quasi-local energy for the Schwarzschild metric using analytic matching in three different coordinate representations and obtained in each case the standard result.  This may give one some confidence in these techniques as well as in the standard value.  On the other hand, as we shall see in the following, there are other coordinate systems in which this analytic technique for determining the reference will lead to other values.

%----------------------------------------------------------------------------------------------------------------------------------------------------------------------
%additional part of Isotropic Schwarzschild ----- END -----
%----------------------------------------------------------------------------------------------------------------------------------------------------------------------

\subsubsection{Eddington-Finkelstein.}
The Eddington-Finkelstein (EF) form of the Schwarzschild metric,
\begin{equation}
\td s^2=-\left(1-\frac{2m}{r}\right)\td\tilde{t}^2-2\varsigma
\frac{2m}{r}\td\tilde{t}\td r+\left(1+\frac{2m}{r}\right)\td r^{2}+r^2\td\Omega^{2},
\label{metric-Eddington}
\end{equation}
(where $\varsigma=-1$ is for incoming and $\varsigma=+1$ is for outgoing and $(t,r,\theta,\varphi)$ is the standard coordinate system of the Schwarzschild metric)
 follows from the time coordinate transformation $\tilde{t}=t-\varsigma 2m\ln(\frac{r}{2m}-1)$,
 which makes the outgoing %%%Liu_110831%%%
 (incoming) radial null geodesics into straight lines of slope $\pm 1$ in the $\tilde{t}-r$ plane.  A principal virtue of this form of the metric is that it is regular at the horizon, $r=2m$.
Rewriting this metric in the ADM form,
\begin{equation}
\td s^2=-\left(1+\frac{2m}{r}\right)^{-1}\td\tilde{t}^2+
\left(1+\frac{2m}{r}\right)\left(\td r-\varsigma \frac{2m/r}{1+2m/r}\td\tilde{t}\right)^2+r^2\td\Omega^{2},
\label{metric-Eddington-ADM}
\end{equation}
leads to the coframe
\begin{eqnarray}
&&\vartheta^{0}=\frac{1}{\sqrt{1+2m/r}}\td\tilde{t},\quad \vartheta^{1}=\sqrt{1+\frac{2m}{r}}\left(\td r-\varsigma\frac{2m/r}{1+2m/r}\td\tilde{t}\right),\nonumber\\
&&\vartheta^{2}=r\td\theta,\quad\vartheta^{3}=r\sin\theta\,\td\varphi.
\end{eqnarray}
%%%Liu_110704%%%
%The associated connection one-form components relevant to our calculation are %%%Liu_110429%%% associate to associated %
%\begin{eqnarray}
%\Gamma^2{}_1=(1+2m/r)^{-1/2}\td\theta,&\quad&
%\Gamma^3{}_1=(1+2m/r)^{-1/2}\sin\theta\,\td\varphi,\nonumber\\
%\Gamma^2{}_0=-\varsigma\frac{2m}{r(1+2m/r)^{1/2}}\td\theta,&\quad&
%\Gamma^3{}_0=-\varsigma\frac{2m}{r(1+2m/r)^{1/2}}\sin\theta\,\td\varphi.%%%Liu_110429%%% missing a \varsigma, also change expression the same as the next equation%
%\end{eqnarray}
%The associated reference values obtained analytically with $m=0$ lead to
%\begin{eqnarray}
%\Delta\Gamma^2{}_1=[(1+2m/r)^{-1/2}-1]\td\theta,&\quad&
%\Delta\Gamma^3{}_1=[(1+2m/r)^{-1/2}-1]\sin\theta\, \td\varphi,\nonumber\\
%\Delta\Gamma^2{}_0=-\varsigma\frac{2m}{r(1+2m/r)^{1/2}}\td\theta,&\quad&
%\Delta\Gamma^3{}_0=-\varsigma\frac{2m}{r(1+2m/r)^{1/2}}\sin\theta\,\td\varphi.\quad%%%Liu_110429%%% missing a \varsigma
%\end{eqnarray}
%
%Then we obtain the corresponding Levi-Civita connection ${\Gamma^{\mu}}_{\nu}$ and the reference variables for ${\overline{\Gamma}^{\mu}}_{\nu}:={\Gamma^{\mu}}_{\nu}(m=0)$, $\bar{\vartheta}^{\mu}:=\vartheta^{\mu}(m=0)$.
%
%%%Liu_110704%%%
The quasi-local energy is obtained by straightforward calculating the corresponding Levi-Civita connection and taking $m=0$ as the reference.
For $\textbf{N}=N^{0}\textbf{e}_{0}+N^{1}\textbf{e}_{1}$ it works out to be
%%%Liu_110704%%%
\begin{eqnarray}
E_{\rm EF}(\textbf{N})=r\left(1-\frac{1}{\sqrt{1+2m/r}}\right)N^{0}-\varsigma\frac{2m}{\sqrt{1+2m/r}}N^{1},
\label{quasi-energy-Edd}
\end{eqnarray}
where $\textbf{e}_{0}=\sqrt{1+2m/r}\partial_{\tilde{t}}+\varsigma \frac{2m}{r\sqrt{1+2m/r}}\partial_{r}$ and $\textbf{e}_{1}=\frac{1}{\sqrt{1+2m/r}}\partial_{r}$.
For the choice of the reference timelike Killing field, $\textbf{N}=\partial_{\tilde{t}}$, this expression yields the value
\begin{equation}
E_{\rm EF}(\partial_{\tilde{t}})=2m-r\left(1-(1+2m/r)^{-1/2}\right),
\end{equation}
with the asymptotic and horizon limits
\begin{eqnarray}
E_{\rm EF}(\partial_{\tilde{t}})_{r\rightarrow\infty}=m,\quad%%%Liu_110718
E_{\rm EF}(\partial_{\tilde{t}})_{r=2m}=\sqrt{2}m.\label{quasi-energy-Edd-ddt}
\end{eqnarray}
For the alternative choice of $\textbf{N}=\textbf{e}_{0}$, the quasi-local energy obtained
from (\ref{quasi-energy-Edd}) is
\begin{equation}
E_{\rm EF}(\textbf{e}_{0})=
r\left(1-(1+2m/r)^{-1/2}\right)=\frac{2m}{\sqrt{1+2m/r}\left(1+\sqrt{1+2m/r}\right)},
\end{equation}
with the asymptotic and horizon limits
\begin{eqnarray}
E_{\rm EF}(\textbf{e}_{0})_{r\rightarrow\infty}=m,\quad%%%Liu_110718
E_{\rm EF}(\textbf{e}_{0})_{r=2m}=m(2-\sqrt{2}).
\end{eqnarray}
From these two time choices we find the curious fact again:
\begin{equation}E_{\rm EF}\left(\frac12(\partial_{\tilde{t}}+\textbf{e}_{0})\right)\equiv m.\end{equation}
Another choice is the unit dual mean curvature vector (outside the horizon)
\begin{eqnarray*}
\hat{\textbf{N}}^{\perp}=\frac{1}{\sqrt{1-4m^2/r^2}}\left(\textbf{e}_{0}-\varsigma\frac{2m}{r}\textbf{e}_{1}\right),
\end{eqnarray*}
the associated quasi-local energy is
\begin{eqnarray}
E_{\rm EF}(\hat{\textbf{N}}^{\perp})&=&
r\left(\frac{1}{\sqrt{1-4m^2/r^2}}-\sqrt{1-\frac{2m}{r}}\right),\\
E_{\rm EF}(\hat{\textbf{N}}^{\perp})_{r\rightarrow\infty}&=&m,\quad%%%Liu_110718
E_{\rm EF}(\hat{\textbf{N}}^{\perp})_{r=2m}\rightarrow\infty.
\label{quasi-energy-Edd-mean}
\end{eqnarray}

\subsubsection{Painlev\'e-Gullstrand.}
Another form of the Schwarzschild metric which is regular at the horizon is the
Painlev\'e-Gullstrand (PG) form:
\begin{eqnarray}
\td s^2&=&-(1-2m/r)\td\tau^{2}-2\varsigma \sqrt{2m/r}\td\tau
\td r+\td r^{2}+r^2\td\Omega^2\nonumber\\
&=&-\td\tau^{2}+\left(\td r-\varsigma\sqrt{2m/r}d\tau\right)^2+r^2\td\Omega^2 \label{PGmetric}
\end{eqnarray}
(where $\varsigma=-1$ means incoming and $\varsigma=+1$ means outgoing). The PG time coordinate is
given by the relation $\td\tau=\td t-\varsigma\frac{\sqrt{2m/r}}{1-2m/r}\td r$. The most noteworthy
feature of this form of the Schwarzschild metric is that the geometry of the spatial %%%Liu_110831%%%
$\tau=\hbox{constant}$ surfaces is {\em flat.}  We choose the coframe
\begin{eqnarray}
\vartheta^{0}=\td\tau,\quad \vartheta^{1}=\td r-\varsigma\sqrt{2m/r}\td\tau,\quad\vartheta^{2}=r\td\theta,\quad\vartheta^{3}=r\sin\theta\,\td\varphi.
\end{eqnarray}
Some of the connection one-form components, namely $\Gamma^2{}_1$, $\Gamma^3{}_1$, $\Gamma^3{}_2$
%%%Liu_110718\begin{equation}
%\Gamma^2{}_1=\td\theta, \quad \Gamma^3{}_1=\sin\theta\,
%\td\varphi, \quad \Gamma^3{}_2=\cos\theta\,\td\varphi,
%%%Liu_110718\end{equation}
have the same values as in Minkowski space (so the corresponding
$\Delta\Gamma$ vanish); whereas the others are
\begin{equation}
\Gamma^1{}_0=-{\varsigma\over2}\sqrt{2m/ r^3}\vartheta^1,\quad
\Gamma^2{}_0=\varsigma\sqrt{2m/ r^3}\vartheta^2,\quad
\Gamma^3{}_0=\varsigma\sqrt{2m/ r^3}\vartheta^3%%%Liu_110720
\end{equation}
(with the corresponding $\Delta\Gamma$ having the same values).

Using the same procedure as above, the quasi-local energy now works out to be
\begin{equation}
E_{\rm PG}(\textbf{N})=-\varsigma\sqrt{2mr}N^{1}.
\end{equation}
For the reference timelike Killing choice, $\textbf{N}=\partial_{\tau}={\bf e}_0-\varsigma\sqrt{2m/r}{\bf e}_1$, this expression yields the value
\begin{eqnarray}
E_{\rm PG}(\partial_{\tau})=2m,\quad\textrm{everywhere.}
\end{eqnarray}
Whereas it is obvious that if $\textbf{N}=\textbf{e}_{0}$, then
\begin{equation}
E_{\rm PG}(\textbf{e}_{0})=0\quad\textrm{everywhere}.
\end{equation}
This latter quasi-local value is consistent with the well-known fact that the  ADM energy vanishes for the PG metric (since the spatial metric of the constant $\tau$ surfaces is just that of flat Euclidean space).
 Once again we find the curious result:
 \begin{equation}E_{\rm PG}\left(\frac12(\partial_{\tau}+\textbf{e}_{0})\right)\equiv m.\end{equation}
On the other hand,  for the unit dual mean curvature vector outside the horizon,
\begin{equation}
\hat{\textbf{N}}^{\perp}=\frac{1}{\sqrt{1-2m/r}}(\textbf{e}_{0}
-\varsigma\sqrt{{2m}/{r}}\textbf{e}_{1}),
\end{equation}
the PG quasi-local energy has the value
\begin{equation}
E_{\rm PG}(\hat{\textbf{N}}^{\perp})=
\frac{2m}{\sqrt{1-2m/r}},
\end{equation}
with the asymptotic and horizon limits
\begin{eqnarray}
E_{\rm PG}(\hat{\textbf{N}}^{\perp})_{r\rightarrow\infty}=2m,\quad%%%Liu_110718
E_{\rm PG}(\hat{\textbf{N}}^{\perp})_{r=2m}\rightarrow\infty.
\end{eqnarray}
Note that this value does not approach the ADM energy at spatial infinity.  It is well known that  that desirable property can only be expected to hold for metrics which fall off faster than $O(r^{-1/2})$, see, e.g.,~\cite{Szabados:2003yn}.

%-----------------------------------------------------------------------------------------------------------------
%          rewritten of this part
%-----------------------------------------------------------------------------------------------------------------

\subsection{FLRW cosmology}

 %%%jmn There seemed to be too much overlap, so I decided to try replaceing all the text cosmological stuff by the cosmological material in the appendix.  And it seemed to me that including a non-vanishing \bar k did not fit well with the other material.

%\subsubsection*{proper radial coordinate}

Now let us consider  dynamic spherically symmetric metrics. The homogeneous-isotropic
 FLRW cosmological metric %%%jmn110722 , %%%Liu_110720
has several equivalent manifestly-isotropic-about-one-point forms with
$\td s^2=-\td t^2+a(t)^2 \td l^2$, where
\begin{eqnarray}
\td l^2&=&\td \chi^2+\Sigma^2(\chi) \td\Omega^2\label{FLRWmetric1}\\
&=& (1-kr^2)^{-1}\td r^2+r^2\td \Omega^2\label{FLRWmetric2}\\
&=& \left[1+kR^2/4\right]^{-2}\left(\td R^2+R^2 \td \Omega^2\right)\label{FLRWmetric3}\\
&=& \left[1+k(x^2+y^2+z^2)/4\right]^{-2}\left( \td x^2+\td y^2+\td z^2\right)\label{FLRWmetric4}.
\end{eqnarray}
The first uses the {\em proper radial coordinate} $\rho=\chi$, with
$\Sigma(\chi)=\{\sinh\chi,\chi,\sin\chi\}$ respectively corresponding to the spatial curvature signature  $k=\{-1,0,+1\}$.

%Here we will take the reference %to be the Minkowski% %%%Liu_110429_FLRW%%%
%space obtained analytically from the respective dynamical metrics by taking $\bar a(t)=1$, $\bar k=0$ $(\bar{\Sigma}=\chi)$, %%%Liu_110429_FLRW%%% add a (\bar{\Sigma})
%and will use the general quasi-local energy expression derived in the \textbf{Appendix}.%%%Liu_110429%%% bold appendix %

Here we will take the reference metric and connection components %%%jmn110501
to be obtained analytically from the respective dynamical ones by taking $\bar a(t)=1$, $\bar k=0$ $(\bar{\Sigma}=\chi)$, %%%Liu_110429_FLRW%%% add a (\bar{\Sigma})
and will use the general quasi-local energy expression derived in the \textbf{Appendix}.%%%Liu_110429%%% bold appendix %

For the first metric form (\ref{FLRWmetric1}), for the quasi-local energy of a sphere at constant $t,\rho$ from (\ref{A:Esph}) with  $A=a(t)$, $B=a(t)\Sigma(\chi)$, we find
\begin{equation}
E_{\rm FLRW}=-a\Sigma\Delta\Sigma',
\end{equation}
which is,
respectively,
\begin{equation}E_{\rm FLRW}=a\{\sinh\chi(1-\cosh\chi),0,\sin\chi(1-\cos\chi)\}.\end{equation}
%
%\subsubsection*{area coordinate}
%
For the {\em area\/} coordinate $\rho=r$, from the metric form (\ref{FLRWmetric2})%%%Liu_110429_FLRW%%% should be eq (66) %
, $A=a(1-kr^2)^{-1/2}$, $B=ar$.
Then from (\ref{A:Esph}) for the quasi-local energy of the sphere at constant $t,r$ we find %%%jmn110722 ,
\begin{equation}
E_{\rm FLRW}=-ar\Delta\sqrt{1-kr^2}=ar(1-\sqrt{1-kr^2}) \label{FLRWareaE}. %%%jmn110722 , and I deleted the following equation, since the information appears below.
\end{equation}
%which is,
% respectively,
%
%\begin{equation} E_{\rm FLRW}=ar\{[1-(1+r^2)^{1/2}],0,[1-(1-r^2)^{1/2}]\}.\label{FLRWareaE}
%\end{equation}
%
%\subsubsection*{isotropic coordinates}
%
For {\em isotropic spherical\/} coordinates take $\rho=R$, and from the metric form (\ref{FLRWmetric3})
$A=a/[1+(k/4)R^2]$, $B=AR$. From (\ref{A:Esph}) for the quasi-local energy of a sphere at constant $t,R$ we find
\begin{equation}
E_{\rm FLRW}=\frac{akR^3}{2[1+(k/4)R^2]^2}. %%%jmn110722 , and I deleted the following equation, since to information appears below.
\end{equation}
%which is, respectively,
%\begin{equation}
%E_{\rm FLRW}={aR^3\over2}\{-(1-R^2/4)^{-2},0,(1+R^2/4)^{-2}\}.
%\end{equation}
We note that the isotropic {\em Cartesian}
formula (\ref{A:Eisc}) with $\Phi=a[1+(k/4)R^2]^{-1}$ obtained from the metric form (\ref{FLRWmetric4}) gives {\em
exactly} this same value. %%%jmn110722 these same values.

%\subsubsection*{energies}

 Although the above results may appear to be different they are in
fact {\em identical}, as can readily be verified using
$\Sigma(\chi)=r=R/(1+kR^2/4)$ with due consideration to the respective ranges of the radial coordinates used in these various representations of the FLRW metric. In summary, for FLRW we have the respective equivalent quasi-local energy values:
\begin{eqnarray}
E_{k=-1}&=&a\sinh\chi(1-\cosh\chi)=ar[1-\sqrt{1+r^2}]={-aR^3\over2(1-R^2/4)^{2}},\\
E_{k=0}&=&0,\\
E_{k=+1}&=&a\sin\chi(1-\cos\chi)=ar[1-\sqrt{1-r^2}]={aR^3\over2(1+R^2/4)^{2}}.
\end{eqnarray}

 It is noteworthy that, according to this measure, the sign of the
quasi-local energy is proportional to $k$, being negative for the open
universe, vanishing for the flat case and positive for the closed
case---but (just as it should be) {\em vanishing} when the whole universe is considered.

These results %%%Liu_110830%%%
(which were first reported in [16]) may be compared with those obtained using the same quasi-local Hamiltonian boundary expression applied
to {\em homogenous} but generally non-isotropic Bianchi cosmological models, using a {\em homogeneous} choice of
reference~\cite{Nester:2008xd}.  That analysis found a vanishing quasi-local value for all Bianchi class A models
(which includes as special cases the isoropic FLRW $k=0$ and $k=+1$ models) and a {\em negative} quasi-local energy
for all class B models (including as a special case the isotropic FLRW %isotropic% %%%Liu_110429%%% one more word: isotropic %
 $k=-1$ model).

It is also noteworthy that in the FLRW $k=-1$ model with vanishing
matter, one finds $a(t)=t$.  It can be directly verified that the
geometry is then really flat Minkowski space, yet our quasi-local Hamiltonian boundary term expression gives a
non-vanishing energy, which, moreover is {\em negative}.  That a negative quasi-local energy value for certain cosmological models can be physically appropriate has been discussed in the work cited in the previous paragraph.  In the present case, the negative quasi-local energy is related to the choice of dynamically expanding comoving observers and their associated choice of reference.  The next section describes an alternative technique for choosing the  reference that will yield a different value for the FLRW quasi-local energy in general and for this curious special case in particular.

%--------------------------------------------------------------------------------------------------------------------
% additional isotropic FLRW -----------END----------------
%--------------------------------------------------------------------------------------------------------------------

%%%jmn110501 Refined the English
%\section*{Remark}
%The trivial choice of reference coframe (information of metric) and connection is setting some specific constant parameters of the dynamic spacetime. For example, Schwarzschild case, $m=0$ is the choice. Under this kind of reference choice, in general, the quasi-local energy depends on the coordinate systems along the same displacement vector (e.g. the dual mean curvature vector). The reason is that this kind of reference choices define different reference connections. For example, let the standard Schwarzschild reference connection denoted by $\overline{\bm{\Gamma}}_{S}$ and Eddington-Finkelstein one by $\overline{\bm{\Gamma}}_{E}$. It is clear that for taking $m=0$ as the reference connection, both of them are anti-symmetric, but for the rule of connection transformation between two \emph{orthonormal coframes}, say $\vartheta^{\mu}_{E}=\Lambda\vartheta^{\mu}_{S}$ (note that it is \emph{dynamic frame}), it should be $\overline{\bm{\Gamma}}_{S}=\td\Lambda\Lambda^{-1}+\Lambda\overline{\bm{\Gamma}}_{E}\Lambda^{-1}$, where $\td\Lambda\Lambda^{-1}$ includes diagonal terms. So for $m=0$ choices, under the orthonormal frame construction, the different reference connections are defined w.r.t different frame choices.

\subsection*{Remark}
The analytic choice of reference coframe and connection was obtained by taking trivial values for some specific constant parameters of the dynamic spacetime.
 For example, for the Schwarzschild case taking $m=0$.
 For this kind of reference choice the quasi-local energy may depend on the coordinate systems along
 the same displacement vector (e.g., $E_{\rm S}(\hat{\textbf{N}}^{\perp})\neq E_{\rm EF}(\hat{\textbf{N}}^{\perp})$). %%%Liu_110502%%%
This can happen because this kind of reference choice may lead to different reference connections. For example, let the standard Schwarzschild reference connection be denoted by $\overline{\bm{\Gamma}}_{\rm S}$ and the Eddington-Finkelstein one by $\overline{\bm{\Gamma}}_{\rm EF}$. It is clear that when taking $m=0$ for the reference connection components, both of them remain anti-symmetric.
%but the rule for the transformation of connection one-forms between two \emph{orthonormal coframes}, say $\bm{\vartheta}_{\rm EF}=\Lambda\bm{\vartheta}_{\rm S}$ (note that this is for the \emph{dynamic frame}), should be $\overline{\bm{\Gamma}}_{\rm S}=\td\Lambda\Lambda^{-1}+\Lambda\overline{\bm{\Gamma}}_{\rm EF}\Lambda^{-1}$, where $\td\Lambda\Lambda^{-1}$, which includes diagonal terms. So the $m\to0$ limits, under the orthonormal frame construction, the different reference connections are defined w.r.t different frame choices.
If the dynamic orthonormal coframes are related by $\bm{\vartheta}_{\rm EF}=\Lambda\bm{\vartheta}_{\rm S}$, then for the reference connection coefficients expressed in the dynamical frames $\overline{\bm{\Gamma}}_{\rm EF} \ne -\td\Lambda\Lambda^{-1}+\Lambda\overline{\bm{\Gamma}}_{\rm S}\Lambda^{-1}$.

%%%jmn110501 This is not yet clear to me.

%%%Liu_110502%%% make some simbol changes

\section{Choice of reference: extremization of energy}

 An alternative strategy for obtaining the reference and displacement vector is via extremization of the quasi-local energy. %%%Liu_110718
   We note that Wang-Yau have used this technique to select a reference for their quasi-local energy expression \cite{Wang:2008jy,WaYa2008}. %%%jmn110722 Here a space has been inserted %%%Liu_110720
   %%%Liu_110718
   This is reasonable in light of the usual desiderata that quasi-local energy should be non-negative and should vanish iff the dynamical space is
   actually flat Minkowski space. %%%Liu_110718%%%Liu_110720
     For if one supposes %%%jmn110722
     that the quasi-local energy expression for any reasonable choice of reference indeed were non-negative, and vanished only if
     the dynamical variables were actually those of Minkowski space, then the quasi-local energy could be expected to have a unique minimum for some
     reference.
     %%%Liu_110718
     Extremizing the energy w.r.t the reference can be viewed as selecting a Minkowski reference that is ``\emph{closest}'' to the dynamical space
     (where the energy value is used to measure how close). %%%Liu_110720
     Here we shall assume that the reference space has a Minkowski \emph{metric} and a connection (which, however, need not necessarily turn out to be flat; technically we will simply keep the \emph{anti-symmetric shape} of the reference connection when it is expressed in the \emph{dynamic orthonormal coframe}).

Since we consider here only spherical symmetric physical spacetimes, the natural choice of the quasi-local two-surface is a constant $t$, $r$ two-sphere;  the tangent space of this surface is expressed by the spherical orthonormal frame basis $\textbf{e}_{2}$, $\textbf{e}_{3}$. This simplifies the boundary expression and also makes it easier to determine an isometric embedding of the two-boundary into the reference space. The extremization comes from extremizing the value of the quasi-local energy over the reference gauge choices, i.e., the extremal value of energy over the reference coordinate transformations. Through the extremizing process, the choice of reference variables and the displacement vector are tied together with the dynamic connection. The displacement vector comes out to be the dual mean curvature vector of the two-boundary. Using this approach, we are able to obtain a quasi-local energy value for the Schwarzschild metric which is independent of the choice of the $t,r$ coordinates and which, moreover, gives zero energy for the $a=t$, $k=-1$ FLRW cosmology, i.e., for the dynamic representation of Minkowski space.

Let us now introduce this process. Suppose the reference metric has the form
\begin{equation}
\td\bar{s}^2=-\td T^2+\td R^{2}+R^2(\td\Theta^{2}+\sin^{2}\Theta \td\Phi^{2}).
\label{emb01}
\end{equation}
To determine the quasi-local energy, we have to obtain the reference connection which is pulled
back from the reference space to the dynamic space via a coordinate transformation. This means
finding $\{T,R,\Theta,\Phi\}$, which are in general functions of $\{t,r,\theta,\varphi\}$. %%%Liu_110830%%%
 Because of the special simplicity of the spherically symmetric metrics, we can assume the coordinate transformation to have the restricted form
\begin{equation}
T=T(t,r),\quad R=R(t,r),\quad \Theta=\theta,\quad \Phi=\varphi.
\label{emb02}
\end{equation}
Then (\ref{emb01}) becomes
\begin{eqnarray}
\td\bar{s}^{2}&=&-(\dot{T}^{2}-\dot{R}^{2})\td t^{2}+2(\dot{R}R'-\dot{T}T')\td t\td r+(R'^{2}-T'^{2})
\td r^{2}+R^{2}\td\Omega^{2}\nonumber\\
&=&\bar{g}_{00}\td t^{2}+2\bar{g}_{01}\td t\td r+\bar{g}_{11}\td r^{2}+\bar{g}_{22}\td\theta^{2}
+\bar{g}_{33}\td\varphi^{2},\label{emb03}
\end{eqnarray}
where $\dot{T}=\partial T/\partial t$, $T'=\partial T/\partial r$,
$\dot{R}=\partial R/\partial t$, $R'=\partial R/\partial r$.
We can  rewrite ({\ref{emb03}}) in the ADM form:
\begin{equation}
\td\bar{s}^{2}=-\frac{\bar{g}_{01}^{2}-\bar{g}_{00}\bar{g}_{11}}{\bar{g}_{11}}\td t^{2}
+\left(
\sqrt{\bar{g}_{11}}\td r +\frac{\bar{g}_{01}}{\sqrt{\bar{g}_{11}}}\td t\right)^{2}+\bar{g}_{22}\td\theta^{2}+\bar{g}_{33}
\td\varphi^{2}.\label{emb04}
\end{equation}
Choose the coframe from (\ref{emb04}):
\begin{eqnarray}
&&\bar{\vartheta}^{0}=\frac{\sqrt{\bar{g}_{01}^{2}-\bar{g}_{00}\bar{g}_{11}}}{\sqrt{\bar{g}_{11}}}\td t,\quad
\bar{\vartheta}^{1}=\sqrt{\bar{g}_{11}}\td r+\frac{\bar{g}_{01}}{\sqrt{\bar{g}_{11}}}\td t
,\nonumber\\
&&\bar{\vartheta}^{2}=\sqrt{\bar{g}_{22}}\td\theta=R\td\theta,\quad
\bar{\vartheta}^{3}=\sqrt{\bar{g}_{33}}\td\varphi=R\sin\theta\,\td\varphi, \label{emb07}
\end{eqnarray}
and the corresponding orthonormal frame is denoted by $\bar{\textbf{e}}_{\mu}$. Next define the connection of the reference:
 \begin{eqnarray}
&&{\overline{\Gamma}^{0}}_{1}={\overline{\Gamma}^{1}}_{0}=\cdots\quad\textrm{not needed below},\nonumber\\
&&{\overline{\Gamma}^{0}}_{2}={\overline{\Gamma}^{2}}_{0}=\bar{P}\td\theta,\quad
{\overline{\Gamma}^{0}}_{3}={\overline{\Gamma}^{3}}_{0}=\bar{P}\sin\theta\,
\td\varphi,\nonumber\\
&&{\overline{\Gamma}^{1}}_{2}=-{\overline{\Gamma}^{2}}_{1}=\bar{Q}\td\theta,\quad
{\overline{\Gamma}^{1}}_{3}=-{\overline{\Gamma}^{3}}_{1}=\bar{Q}\sin\theta\,
\td\varphi,\nonumber\\
&&{\overline{\Gamma}^{2}}_{3}=-{\overline{\Gamma}^{3}}_{2}=-\cos\theta\, \td\varphi,\label{3.97}
\end{eqnarray}
where the explicit form of the component functions are
\begin{eqnarray}
\bar{P} &=& - \frac{\bar{g}_{01}R'-\bar{g}_{11}\dot{R}}{\sqrt{\bar{g}_{11}}\sqrt{\bar{g}_{01}^{2}-\bar{g}_{00}\bar{g}_{11}}} = \frac{\pm T'}{\sqrt{R'^{2}-T'^{2}}}, \label{3.98}
\\
\bar{Q} &=& -\frac{R'}{\sqrt{\bar{g}_{11}}} = \frac{-R'}{\sqrt{R'^{2}-T'^{2}}}.\label{3.99}
\end{eqnarray}
%\begin{equation}
%\bar{P}=\frac{\pm T'}{\sqrt{R'^{2}-T'^{2}}},\quad\bar{Q}=\frac{-R'}{\sqrt{R'^{2}-T'^{2}}}.\label{3.99}
%\end{equation}

The dynamic connection coefficients have a similar form.  The ones that we will explicitly need can likewise be parameterized by two functions:
 \begin{eqnarray}
&&{{\Gamma}^{0}}_{2}={{\Gamma}^{2}}_{0}={P}\td\theta,\quad
{{\Gamma}^{0}}_{3}={{\Gamma}^{3}}_{0}={P}\sin\theta\,
\td\varphi,\nonumber\\
&&{{\Gamma}^{1}}_{2}=-{{\Gamma}^{2}}_{1}={Q}\td\theta,\quad
{{\Gamma}^{1}}_{3}=-{{\Gamma}^{3}}_{1}={Q}\sin\theta\,
\td\varphi.\label{connection-extrem-dynamic}
\end{eqnarray}

We assume that the displacement vector $N=N^{0}\textbf{e}_{0}+N^{1}\textbf{e}_{1}$ is in the normal plane of the constant $t$, $r$ surface,
 where $N^{0}$, $N^{1}$ are functions of $(t,r)$ \emph{only},
 %%%Liu_110719
 independent of $T'$ and $R'$.
 %%%Liu_110714%%%
  The second term of (\ref{cov-bound-GR01}) is not vanishing in general. Considering the spherical symmetric case, the boundary integral over the constant $(t,r)$ surface $\cal{S}$ involves the $\Delta\eta_{01}$ term:
\begin{eqnarray}
I&=&\frac{1}{\kappa}\oint(\bar{D}^{1}\bar{N}^{0}-\bar{D}^{0}\bar{N}^{1})(g_{22}-\bar{g}_{22})\sin\theta \textrm{d}\theta\wedge\textrm{d}\varphi\nonumber\\
&=&\frac{1}{2}(\bar{D}^{1}\bar{N}^{0}-\bar{D}^{0}\bar{N}^{1})(g_{22}-\bar{g}_{22})|_{_{\cal{S}}},\label{second term of boundary}
\end{eqnarray}
where $\bar{N}^{\mu}$ is the component expressed in the holonomic basis of reference: $\textbf{N}=N^{0}\textbf{e}_{0}+N^{1}\textbf{e}_{1}=\bar{N}^{0}\partial_{T}+
\bar{N}^{1}\partial_{R}$.

The quasi-local energy works out to be
\begin{eqnarray}
\hspace{-1cm} E&=&\frac{1}{16\pi}\oint\mathcal{B}=\frac{1}{16\pi}\oint
\Delta{\Gamma^{\alpha}}_{\beta}\wedge\iota_{N}{\eta_{\alpha}}^{\beta}+I\nonumber\\
\hspace{-1cm} &=&\frac{1}{4\pi}\oint\left[\left(Q+\frac{R'}{\sqrt{R'^{2}-T'^{2}}}\right)\frac{N^{0}}{\sqrt{g_{22}}}-
\left(P\mp\frac{T'}{\sqrt{R'^{2}-T'^{2}}}\right)\frac{N^{1}}{\sqrt{g_{22}}}\right]
\vartheta^{2}\wedge\vartheta^{3}+I\nonumber\\
\hspace{-1cm} &=&\sqrt{g_{22}}\left(Q+\frac{R'}{\sqrt{R'^{2}-T'^{2}}}\right)N^{0}-\sqrt{g_{22}}\left(P\mp\frac{T'}{\sqrt{R'^{2}-T'^{2}}}\right)N^{1}+I.
\label{general-energy}
\end{eqnarray}

In the previous section, the choices of reference variables for Schwarzschild were obtained by taking $m=0$.
The term $I$ in (\ref{general-energy}) vanishes because $(g_{22}-\bar{g}_{22})|_{_{\cal{S}}}=0$. %%%Liu_110715%%%
 In this section, we will let the functions $T(t,r)$ and $R(t,r)$ be undetermined, and through the extremization of energy find out what these functions should be. However we will \emph{require} $R(t,r)|_{_{\cal{S}}}=\sqrt{g_{22}}$. This makes the two-surfaces in the dynamic spacetime and the reference space isometric. Since the Hamiltonian boundary term is a quantity dependent on the \emph{quasi-local two-surface}, the isometric requirement of the two-surface is reasonable, and furthermore, it simplifies the boundary expression by making $I=0$.

%%%Liu_110714%%%
% rewrite this part %

For any given fixed $\bf N$, extremize %%%jmn110722 Extremize
the energy by requiring the vanishing of the partial derivative with respect to $T'$
(it is easy to check that taking the partial derivative of (\ref{general-energy}) w.r.t.\ %%%jmn110722
$R'$ gives the same condition):
\begin{eqnarray}
\frac{\partial E}{\partial T'}=\sqrt{g_{22}}\frac{R'(T'N^{0}\pm R'N^{1})}{(R'^{2}-T'^{2})^{3/2}}=0,\quad \Rightarrow T'=\mp\frac{N^{1}}{N^{0}}R'.\label{mini-dEdT}
\end{eqnarray}
Substitute into (\ref{general-energy}) then
\begin{eqnarray}
E=\sqrt{g_{22}}[(Q+N^{0})N^{0}-(P+N^{1})N^{1}].\label{quasi-energy-QN0+PN1}
\end{eqnarray}
Suppose we choose the normalized time-like displacement $\textbf{N}$ which means $-(N^{0})^{2}+(N^{1})^{2}=-1$, then (\ref{quasi-energy-QN0+PN1}) becomes
\begin{eqnarray}
E=\sqrt{g_{22}}(1+QN^{0}-PN^{1}).\label{quasi-energy_1+QN0-PN1}
\end{eqnarray}
This result implies that the quasi-local energy depends on the \emph{free choice} of $\textbf{N}$.
We further look at the extremal value w.r.t all the displacements.
Let $N^{0}=\cosh\alpha$, $N^{1}=\sinh\alpha$, take the extremization of the %%%Liu_110720
quasi-local energy value (\ref{quasi-energy_1+QN0-PN1}), %%%jmn110722
\begin{eqnarray}
\frac{\partial E}{\partial\alpha}=\sqrt{g_{22}}(Q\sinh\alpha-P\cosh\alpha)=0,\quad\Rightarrow\frac{\sinh\alpha}{\cosh\alpha}=\frac{N^{1}}{N^{0}}=\frac{P}{Q},
\end{eqnarray}
%\begin{eqnarray}
%&&\frac{\partial E}{\partial T'}=\sqrt{g_{22}}\frac{R'(T'N^{0}\pm
%R'N^{1})}{(R'^{2}-T'^{2})^{3/2}}\nonumber\\
%&&+\sqrt{g_{22}}\left[\left(Q+\frac{R'}{\sqrt{R'^{2}-T'^{2}}}\right)
%\frac{\partial N^{0}}{\partial
%T'}-\left(P\mp\frac{T'}{\sqrt{R'^{2}-T'^{2}}}\right) \frac{\partial
%N^{1}}{\partial T'}\right]=0.\label{mini-dEdT}
%\end{eqnarray}
%Choose the \emph{unit timelike} displacement vector \textbf{N},
%that means $-(N^{0})^{2}+(N^{1})^{2}=-1$. Now $N^{0}$
%and $N^{1}$ are functions of $(t,r)$, and $\frac{\partial
%N^{1}}{\partial T'}$ is related to $\frac{\partial
%N^{0}}{\partial T'}$. It is important to note that on the 2-surface of constant $t,r$ the values of the quantities $N^{0}$ and
%$\frac{\partial N^{0}}{\partial T'}$ are independent. Hence
%we can require that each part of (\ref{mini-dEdT}) vanishes separately. From the
%vanishing first part, we get $T'N^{0}\pm R'N^{1}=0$. For
%$R'\geq0$, future directed \textbf{N} means positive $N^{0}$; hence
%\begin{equation}
%N^{0}=\frac{R'}{\sqrt{R'^{2}-T'^{2}}},\quad N^{1}=\frac{\mp
%T'}{\sqrt{R'^{2}-T'^{2}}}.\label{disp-comp-mini-e}
%\end{equation}
%Note that $R'^{2}>T'^{2}$. Substitute these results into (\ref{mini-dEdT}); the
%second part gives the condition
%\begin{equation}
%\sqrt{g_{22}}\frac{ R'}{(R'^{2}-T'^{2})^{3/2}}(QT'\pm PR')=0.
%\end{equation}
%
%
%
% end of the rewritten %%%Liu_110714%%%
then we have the relation
\begin{equation}
T'=\mp\frac{P}{Q}R',\label{T-R}
\end{equation}
and
\begin{eqnarray}
N^{0}=\frac{1}{\sqrt{1-P^{2}/Q^{2}}},\quad N^{1}=\frac{P}{Q\sqrt{1-P^{2}/Q^{2}}}.\label{m0m1-final}
\end{eqnarray}
%%%Liu_110714%%%
%  delete a sentence
%%%
Consequently, the quasi-local energy (\ref{general-energy}) has the extreme value
\begin{equation}
E_{\rm ex}=\sqrt{g_{22}}(1+Q\sqrt{1-{P^{2}}/{Q^{2}}}).\label{mini-energy-semi}
\end{equation}
We can see that $P$, $Q$ are determined purely by the metric of the dynamic spacetime. There is no longer any information of the
reference frame or the displacement vector in this energy expression. %But
%the relation (\ref{T-R}) should be satisfied, which restricts the choice of reference to (\ref{restrict-reference}), and
%(\ref{m0m1-final}) gives the displacement vector.
With the vectors $\{\textbf{e}_{2},\textbf{e}_{3}\}$  tangent to the two-surface,
the dual mean curvature vector of the two-surface in the dynamic space is
$\textbf{N}^{\bot}=-k\textbf{e}_{0}+p\textbf{e}_{1}$, where $k=2Q/\sqrt{g_{22}}$ is
the extrinsic curvature w.r.t.\ %%%jmn110722
 the space-like normal $\textbf{e}_{1}$ and $p=-2P/\sqrt{g_{22}}$ is the extrinsic curvature w.r.t.\ %%%jmn110722
 the time-like normal $\textbf{e}_{0}$. Then the vector $\hat{\textbf{N}}^{\bot}:=\textbf{N}^{\bot}/|\textbf{N}^{\bot}|$ has the components (\ref{m0m1-final}),
%%%Liu_110704%%%
where $|\textbf{N}^{\bot}|:=\sqrt{-\langle\textbf{N}^{\bot},\textbf{N}^{\bot}\rangle}$.
%%%Liu_110704%%%

Rewrite (\ref{mini-energy-semi}) by replacing $Q$ and $P$ with the extrinsic curvature $k$ and $p$: %%%jmn110722 , which is
\begin{eqnarray}
E_{\rm ex}=\frac{g_{22}}{2}\left(2/\sqrt{g_{22}}+(-k)\sqrt{1-p^{2}/k^{2}}\right).
\end{eqnarray}
Here we use the definition of extrinsic curvature which is
$k_{ab}:=-\langle\nabla_{a}\textbf{e}_{1},\textbf{e}_{b}\rangle$, and $p_{ab}:=-\langle\nabla_{a}\textbf{e}_{0},
\textbf{e}_{b}\rangle$, $a,b=2,3$; the trace is $k=\delta^{ab}k_{ab}$ and $p=\delta^{ab}p_{ab}$.
By this convention $k$ is \emph{negative} and $\textbf{N}^{\perp}$ is \emph{time-like} for the
\emph{dual mean curvature vector}, so that $(-k)\sqrt{1-p^{2}/k^{2}}=\sqrt{k^{2}-p^{2}}=|\textbf{N}^{\perp}|$.
%%%Liu_110720
 Equation (\ref{connection-extrem-dynamic}) %%%Liu_110721
 implies that the trace of the reference extrinsic
 curvature $k_{0}$ is given %%%Liu_110720
 by taking $Q=-1$, so that $k_{0}=-2/\sqrt{g_{22}}$. Consequently,%%%Liu_110429%%% I did some revised about the sign problem%
\begin{eqnarray}
E_{\rm ex}=\frac{g_{22}}{2}\left(|\textbf{N}^{\perp}|-k_{0}\right),\label{quasi-energy-mean-ex}
\end{eqnarray}
which is the same as the Liu-Yau result~\cite{Liu:2003bx}.
%%%Liu_110719
%(the factor $2$ comes from the different expression of quasi-local energy, e.g. the factor $1/\kappa$ in Liu-Yau's paper and $1/2\kappa$ we use here).
% there should be no factor 2 problem
%Liu_110719
Now let us check the following cases.

\subsection{Standard Schwarzschild.}

The functions necessary here are found from (\ref{sch-met},\ref{connection-Sch}) to be
\begin{equation}
\sqrt{g_{22}}=r,\quad P=0,\quad Q=-\sqrt{1-2m/r}.\nonumber
\end{equation}
With these expressions, the extreme energy (\ref{mini-energy-semi}) %%%Liu_110429%%% correct the wrong citation of equation %
  works out to have the standard Brown-York value:
\begin{equation}
E_{\rm exS}=r(1-\sqrt{1-2m/r}).
\end{equation}
In this case, as $P=0$ we have $T'=0$ and the displacement vector
$\textbf{N}$ is equal to $\textbf{e}_{0}$. The reference here
could be found from
$T=T(t)$, $R=R(t,r)$, $\Theta=\theta$, $\Phi=\varphi$, with the restriction $R(t,r)|_{_{\cal{S}}}=r$.

\subsection{Eddington-Finkelstein.}
The necessary functions obtained from (\ref{metric-Eddington}) now have the values
\begin{equation}
\sqrt{g_{22}}=r,\quad P=\varsigma\frac{2m/r}{\sqrt{1+2m/r}},\quad
Q=-\frac{1}{\sqrt{1+2m/r}}.  \nonumber
\end{equation}
With these expressions the extreme energy (\ref{mini-energy-semi}) %%%Liu_110429%%% correct the wrong citation of equation %
again comes out to be
\begin{equation}
E_{\rm exEF}=r(1-\sqrt{1-2m/r}),
\end{equation}
which is again the standard value.
The condition which restricts the choice of reference is
\begin{equation}
T'=-\frac{P}{Q}R'=\varsigma\frac{2m}{r}R'.
\end{equation}
Then one can set any function $R(t,r)$ with the restriction $R|_{_{\cal{S}}}=r$, and solve for
$T(t,r)$. The displacement vector can also be determined from
(\ref{m0m1-final}) to be \begin{equation}\textbf{N}=(\textbf{e}_{0}-\varsigma\frac{2m}{r}\textbf{e}_{1})/\sqrt{1-4m^{2}/r^{2}},\end{equation}
which is the dual mean curvature vector.

\subsection{Painlev\'e-Gullstrand.}
The necessary functions found from (\ref{PGmetric}) are now
\begin{equation}
\sqrt{g_{22}}=r,\quad P=\varsigma\sqrt{2m/r},\quad Q=-1 . \nonumber
\end{equation}
Using these expressions the extreme energy (\ref{mini-energy-semi}) works out to be %%%Liu_110429%%% correct the wrong citation of equation %
\begin{equation}
E_{\rm exPG}=r(1-\sqrt{1-2m/r}).
\end{equation}
Thus once again we found it to have the standard value.
The condition which restricts the choice of reference is
\begin{equation}
T'=-\frac{P}{Q}R'=\varsigma\sqrt{\frac{2m}{r}}R'.
\end{equation}
The displacement vector is $\textbf{N}=(\textbf{e}_{0}-\varsigma\sqrt{2m/r}\textbf{e}_{1})/\sqrt{1-2m/r}$.

In a similar fashion, if one considers the functions $\sqrt{g_{22}}$, $P$, and $Q$ associated with the {\em spherical isotropic} Schwarzschild coframe (\ref{Schw-sph-iso-coframe}), one will once again obtain from (\ref{mini-energy-semi}) the standard quasi-local energy value.

\subsection{FLRW cosmology}

For the dynamic FLRW cosmological models, from the extreme energy expressions (\ref{mini-energy-semi}) it is sufficient to calculate the quasi-local energy using the metric
functions $\sqrt{g_{22}}=a(t)r$, $P=\dot{a}r$, $Q=-\sqrt{1-kr^{2}}$ obtained from (\ref{FLRWmetric1})---the other forms of the FLRW metric would lead to the same answer.  The result is
\begin{equation}
E_{\rm exFLRW}=ar(1-\sqrt{1-kr^{2}-\dot{a}^{2}r^{2}})=\frac{ar^3[k+\dot a{}^2]}{1+\sqrt{1-kr^{2}-\dot{a}^{2}r^{2}}}.\label{exFLRW}
\end{equation}
In contrast to the analytic result (\ref{FLRWareaE}), which can be {\em negative}, with the aid of the Friedmann cosmological equation,
\begin{equation}
\frac{{\dot a}{}^2}{a^2}+\frac{k}{a^2}=\frac{8\pi}{3}\rho,
\end{equation}
it can be seen that this value is {\em non-negative}.  There is no contradiction, the present quasi-local energy value (\ref{exFLRW}) corresponds to {\em non-expanding} observers using a different reference.

Consider in particular the special test case $k=-1$, $a(t)=t$. We find the energy value: $E=0$. We can find a simple reference choice $R=tr$ (actually, this needs be satisfied only on the two-boundary, not necessary in the whole space), then from (\ref{T-R}) we obtain $T=
t\sqrt{1+r^{2}}$. %%%Liu_110502%%% change sign to be positive reference time
It is well-known that the 4-geometry
\begin{equation}
-\td t^2+\frac{t^2}{1+r^2}\td r^2+t^{2}r^2\td \Omega^2%%%Liu_110429%%% an error %
\end{equation}
is actually Minkowski
space, and \emph{zero energy} is just the value we expect.

\subsection{Discussion}
In contrast to the analytic approach, here we used energy extremization to select the reference and displacement vector field.  We tested the resulting quasi-local energy expression on several forms of the Schwarzschild metric, obtaining in each case the standard quasi-local energy value.  We also tested the expression on the FLRW cosmological metric, obtaining a new result for the FLRW quasi-local energy. In both cases the time displacement vector field turns out to be the dual mean curvature vector, and the quasi-local energy value has the desirable property of being {\em non-negative}, vanishing iff the dynamic geometry is flat Minkowski space.

\section{Conclusion}

The covariant Hamiltonian formalism has been incomplete in one aspect.  The Hamiltonian boundary term, whose value determined the quasi-local quantities, in addition to depending on the dynamical fields, also necessarily depends on  reference values for these dynamical fields (which specify the ground state with vanishing quasi-local quantities) along with a spacetime displacement vector field.  However, no specific proposal had been made as to how to choose these unspecified quantities.   Here, for Einstein's GR, for certain spherically metrics (the static Schwarzschild metric and the dynamical homogeneous isotropic cosmologies), following~\cite{Liuthes} we have considered two techniques for choosing the reference and vector field.

The first (which goes back to~\cite{Chen:1998aw,CMthesis}) depends on an analytic choice of reference fields obtained by taking trivial values for certain
parameters in the metric and connection coefficients.  For the usual time slicing for several spatial metrics this leads to the standard Brown-York
quasi-local energy for the Schwarzschild metric, %%%Liu_110720
but to different energy expressions for the alternative time slicings of the Eddington-Finkelstein and Painlav\'e-Gulstrand metrics.  For the FLRW cosmological metrics it leads to a quasi-local energy value which is proportional to the sign of the spatial curvature (and thus a negative energy for a certain dynamical slicing of Minkowski space).

The other approach uses extremization of the quasi-local energy to select an optimal reference and timelike vector.  The resulting quasi-local energy for these spherically symmetric metrics is independent of the coordinates and is {\em non-negative} (for both the Schwarszchild and FLRW metrics) and vanishes only for Minkowski space.  For the Schwarzchild metric it gives the standard quasi-local value.
Going beyond the present work, there have been further developments in the energy optimization approach;  a brief letter describing this has already appeared~\cite{Chen:2009zd}.

\appendix

\section{General formulas for spherical analytical reference choice}

Here we briefly present the quasi-local energy calculations using an analytic reference choice for general spherically symmetric metrics in several different coordinates and coframes.
\subsection*{Spherical frames}
Consider orthonormal co-frames using spherical coordinates of the form:
\begin{equation}
\vartheta^\tau:=N\,\td\tau, \quad \vartheta^\rho:=A\, \td\rho, \quad
\vartheta^\theta:=B\,\td\theta, \quad \vartheta^\varphi:=B \sin\theta
\,\td\varphi,\label{A:S-coframe}
\end{equation}
where $N, A, B$ are functions only of the general time and radial coordinates  $\tau,\rho$.
The (metric-compatible hence anti-symmetric) connection one-form coefficients are readily obtained from the differentials
\begin{eqnarray}
\td\vartheta^\tau&=&(N'/NA)\vartheta^\rho\wedge\vartheta^\tau,\\
\td\vartheta^\rho&=&(\dot A/AN) \vartheta^\tau\wedge \vartheta^\rho,\\
\td\vartheta^\theta&=&(\dot B/BN) \vartheta^\tau\wedge
\vartheta^\theta+(B'/AB)
\vartheta^\rho\wedge \vartheta^\theta,\\
\td\vartheta^\varphi&=&(\dot B/BN) \vartheta^\tau\wedge
\vartheta^\varphi+(B'/AB) \vartheta^\rho\wedge
\vartheta^\varphi+(1/B)\cot\theta\,
\vartheta^\theta\wedge\vartheta^\varphi,
\end{eqnarray}
where dot and prime represent respectively the $\tau$ and $\rho$ partial derivatives.
The ones of particular interest to us are
\begin{equation}
\Gamma^\theta{}_\rho = \frac{B'}{AB} \vartheta^\theta = \frac{B'}{A} \td\theta, \quad
\Gamma^\varphi{}_\rho = \frac{B'}{AB} \vartheta^\varphi = \frac{B'}{A} \sin\theta\,
\td\varphi.\label{A:S-conn}
\end{equation}
We will also need the associated reference values,
\begin{equation}
\bar\Gamma^\theta{}_\rho=(\bar B'/\bar A)\,\td\theta, \quad
\bar\Gamma^\varphi{}_\rho=(\bar B'/\bar A)\sin\theta\, \td\varphi,\label{A:S-refconn}
\end{equation}
 which we here have {\it
assumed} to be given {\em analytically} (by taking limits like $m\to0$) and thereby affecting the transformations $A\to \bar A$, $B\to \bar B$.  In our calculation we will need
\begin{equation}
\Delta\Gamma^\theta{}_\rho=\Delta(B'/ A)\,\td\theta, \quad
\Delta\Gamma^\varphi{}_\rho=\Delta( B'/ A)\sin\theta\, \td\varphi.\label{A:deltagamma}
\end{equation}

\subsection*{Cartesian frame and coordinates}
Spherically symmetric metrics may be also be described by Cartesian coframes using Cartesian spatial coordinates (labeled by latin indies with range 1,2,3) in the form
\begin{equation}\vartheta^\tau=N\, \td\tau, \qquad
\vartheta^k=\Phi\, \td x^k,\label{A:C-coframe}
\end{equation}
where $N,\Phi$ are functions of $\tau$, $R$ with $R^2=x^k x_k$. We
find
\begin{eqnarray}
\td\vartheta^\tau&=&(N'/N\Phi R)x_k \vartheta^k\wedge \vartheta^\tau,\\
\td\vartheta^k&=&({\dot\Phi}/N\Phi)\vartheta^\tau\wedge\vartheta^k+(\Phi'/\Phi^2
R)x_m \vartheta^m\wedge \vartheta^k.
\end{eqnarray}
The connection coefficients of particular interest are found to be
\begin{equation}
\Gamma^{ij}=(\Phi'/\Phi^2 R)(x^j\vartheta^i-x^i\vartheta^j).\label{A:C-conn}
\end{equation}
We here {\em assume} that the associated reference is given by the Minkowski space obtained by {\em analytically} restricting
these formulas to $N=1$, $\Phi=1$; thus the reference connection is
such that its values {\em vanish}. This is just what we expect for the
Minkowski frame determined by the coordinates $x^\mu$.  Since the reference connection vanishes the Cartesian case provides a good check for our calculations.

Our choice of particular non-vanishing reference
values for the various spherical representations can be understood
as just what is needed, as we shall see, to arrange to give the same
results in all of these frames.

\subsection*{Energy expression}

We are interested in the particular
quasi-local energy given by our preferred Hamiltonian boundary term 2-form expression (\ref{cov-bound-GR01})  with vanishing 2nd term:
\begin{equation}
2\kappa{\cal B}({\bf N}):=\Delta \Gamma^\alpha{}_\beta\wedge
\iota_{\textbf{\tiny N}}\eta_\alpha{}^\beta %+({\bar D} N)\Delta\eta
.\label{A:boundaryterm}
\end{equation}
Here we will take $\textbf{N}$ to be
the unit time-like displacement on the boundary, which is at
constant $\rho,\tau$.  The other choices considered in the text are proportional to this choice
For {\em spherical frames}
with the displacement {\it choice} $\textbf{N}=\textbf{e}_\perp$ %%%Liu_110429%%% bold N and e %
(i.e., one unit of
proper time orthogonal to the constant ``time'' hypersurface), our Hamiltonian boundary term quasi-local
energy 2-form expression reduces to
\begin{eqnarray}
2\kappa{\cal B}(\textbf{e}_\bot)&=&\Delta\Gamma^{ab}\wedge\eta_{\bot a %%%Liu_110429%%% bold e %
b}=2 \Delta\Gamma^{\rho
\theta}\wedge\eta_{\bot\rho\theta}+2\Delta\Gamma^{\rho
\varphi}\wedge\eta_{\bot\rho\varphi}\nonumber\\
&=&4\Delta\Gamma^{\rho
\theta}\wedge\eta_{\bot\rho\theta}= 4\Delta\Gamma^{\rho
\theta}\wedge\vartheta^\varphi=-4B\Delta (B'/A)\td\Omega.
\end{eqnarray}
The associated quasi-local energy, obtained by integrating over a 2-sphere at constant $\tau,\rho$ (with $\kappa=8\pi$), has the value
\begin{equation}
E_{\rm S}(\textbf{e}_\perp)=-B\Delta (B'/A).\label{A:Esph}%%%Liu_110429%%% bold e %
\end{equation}
It is notable that the result is not explicitly dependent on the {\em lapse}  $N$.

On the other hand, for the {\em Cartesian frame} we find
\begin{eqnarray}
2\kappa{\cal B}(\textbf{e}_\bot)%%%Liu_110429%%% bold e %
&=&\Delta\Gamma^{ij}\wedge\eta_{\bot
ij}= 2 (\Phi'/\Phi^2 R)x^j\vartheta^i\wedge\eta_{\bot
ij}\nonumber\\
&=&-4(\Phi'/\Phi^2 R)x^k\eta_{\bot k}= -4\Phi' R^2\td\Omega.
\end{eqnarray}
Hence, in this case, the quasi-local energy obtained from integration over the 2-sphere at constant $\tau,R$ is given simply by
\begin{equation}
E_{\rm C}(\textbf{e}_\perp)=-R^2\Phi'.\label{A:Eisc}%%%Liu_110429%%% bold e %
\end{equation}

%\subsubsection*{momentum}
%Let us also look at the ``momentum'', by spherical symmetry the only
%significant component in the spherical frames is
%\begin{equation}
%{\cal B}(e_\rho)=\Delta\Gamma^{\alpha\beta}\wedge\eta_{\alpha\beta
%\rho}=4\Gamma^{\tau\theta}\wedge\eta_{\tau\theta\rho}=-4\frac{\dot
%B}{N} d\theta \wedge \vartheta^\varphi=-4\frac{B\dot B}{N} d\Omega,
%\end{equation}
%leading to
%\begin{equation}
%p_\rho=-B{\dot B}/N.
%\end{equation}

%In the isotropic cartesian representation we get
%\begin{equation}
%p_\rho=-R^2\Phi {\dot \Phi}/N.
%\end{equation}

%\medskip
\subsection*{Schwarzschild
Applications}
%\medskip

In particular we have
%
%\subsubsection*{area coordinate}
%
for the {\em area\/} coordinate from (\ref{Standard Sch coframe}) $A=(1-2m/r)^{-1/2}$, $B=r$. Using these in the general spherical energy expression (\ref{A:Esph}) gives
 \begin{equation} E_{\rm S}(\textbf{e}_\perp)=r[1-(1-2m/r)^{1/2}],\label{A:Eschws}%%%Liu_110429%%% bold e %
 \end{equation}
which is the {\em standard\/} energy value (\ref{energy-Sch-e0}).
An equivalent expression,
\begin{equation}
E_{\rm S}(\textbf{e}_\perp)=\frac{2m}{1+\sqrt{1-2m/r}},%%%Liu_110429%%% bold e %
\end{equation}
more clearly reveals the horizon and asymptotic limits.
%
%\subsubsection*{isotropic spherical coordinate}
%
For {\em isotropic spherical\/} coordinates, from (\ref{Schw-sph-iso-coframe}), $A=(1+m/2R)^2$, $B=R(1+m/2R)^2$.  Using these in the general spherical expression (\ref{A:Esph}) yields the quasi-local energy
\begin{equation}E_{\rm S}(\textbf{e}_\perp)=m(1+m/2R).\label{A:Eschwi}%%%Liu_110429%%% bold e %
\end{equation}
Recalling that $r=R(1+m/2R)^2$, we find that this is actually the same as the {\em standard\/} value
(\ref{A:Eschws}).
%
%\subsubsection*{Isotropic Cartesian}
On the other hand for the {\em isotropic Cartesian\/} frame, using from (\ref{Schw-cart-iso}),
$\Phi=(1+m/2R)^2$ in the Cartesian expression (\ref{A:Eisc}), turns out to give
%\begin{equation}
%E=-R^2\Phi'=-R^2 2(1+m/2R)(-m/2R^2)=m(1+m/2R)\label{A:EiscC}
%\end{equation}
the same value (\ref{A:Eschwi}),  equivalent to the {\em standard\/} value (\ref{A:Eschws}).  As this case has a vanishing reference connection, it provides an important confirmation, not only for the standard energy value, but also especially for our analytic technique of choosing the the reference.

\section*{Acknowledgement}
This work was supported by the National
Science Council of the R.O.C. under the grants NSC-99-2112-M-008-004
(JMN) and NSC 99-2112-M-008-005-MY3 (CMC) and in part by the
National Center of Theoretical Sciences (NCTS). Special thanks for Ming-Fan Wu, whose ideas were very helpful and made this article more complete and satisfactory.

\end{document}